\documentclass[single]{cambridge6Atight}
\usepackage{aas_macros}
\usepackage{natbib}
\usepackage{rotating}
\usepackage{floatpag}
\rotfloatpagestyle{empty}
\usepackage{graphicx}
\usepackage{multind}
\begin{document}


\chapter{5.6.- Stellar dynamics: rotation, convection, and magnetic fields}
\begin{center}
{\normalsize Savita Mathur$^1$, Jer\^ome Ballot$^{2,3}$ \& Rafael A. Garc\'\i a$^4$\\
\vspace{0.3cm}
$^1$Space Science Institute, 4750 Walnut Street, Suite 205, Boulder, Colorado 80301, USA\\
$^2$CNRS, Institut de Recherche en Astrophysique et Plan\'etologie, 14 avenue Edouard Belin, 31400 Toulouse, France\\
$^3$Universit\'e de Toulouse, UPS-OMP, IRAP, 31400 Toulouse, France
$^4$Laboratoire AIM, CEA/DSM -- CNRS - Univ. Paris Diderot -- IRFU/SAp, Centre de Saclay, 91191 Gif-sur-Yvette Cedex, France\\
}
\end{center}

\section{Introduction}
Stars are changing entities in a constant evolution during their lives. At non-secular time scales --from seconds to years-- the effect of dynamical processes such as convection, rotation, and magnetic fields can modify the stellar oscillations. Convection excites acoustic modes in solar-like stars, while  rotation and  magnetic fields can perturb the oscillation frequencies lifting the degeneracy in the azimuthal component $m$ of the eigenfrequencies (see Chapter 3.2 for the case in which rotation is { slow} and first order perturbative theory can be used). Moreover, the interaction between rotation, convection, and magnetic fields can produce magnetic dynamos, which sometimes yield to regular magnetic activity cycles. 

In this chapter we review {how stellar dynamics can be studied and explain what long-term seismic observations can bring to the understanding of this field.} Thus, we show how we can study some properties of the convective time scales operating in a star like the Sun. We also compare the stratified information we can obtain on the internal (radial) differential rotation from main sequence solar-like stars, to the Sun, and to more evolved sub giants, and giants. 
We complement this information on the internal rotation with the determination of the surface (latitudinal differential) rotation obtained directly from the light curves. Indeed, when stars are active there can be spots on their surfaces dimming the light emitted. When the star rotates, the emitted light will be modulated by the presence of these spots with a period corresponding to the rotation rate at the active latitudes (where the spots develop). { We finally give a brief summary of stellar magnetic studies based on spectroscopic observations and then we} discuss the use of seismology to better understand the stellar magnetism of solar-like stars and the existence of possible magnetic cycles. 
We conclude this chapter by discussing the seismology of fast rotating stars and, from a theoretical point of view, what are the current challenges to infer properties of the internal structure and dynamics of intermediate- and high-mass stars.

For stars harboring planets, activity is an important factor to define the habitable zone --at least for carbon-based life-- as it is known on Earth. If a star is very active, the energetic particles reaching the planets located in the habitable zone can destroy the elements necessary for the development of life unless the planets develop a protection { like the Earth magnetosphere preserving our planet} from these particles. Therefore, stronger constraints are imposed to the planets in order to be good candidates for the development of life.

Understanding those magnetic activity cycles is a necessity to allow us to properly predict them. Unfortunately, we are far from a complete understanding of the physical processes responsible for these magnetic cycles. For the first time in history, we are able to access the internal structure of many stars while putting, at the same time, observational constraints on the convection properties, rotation, and magnetic fields for stars other than the Sun in many different configurations of the parameter space. Asteroseismology is changing our preconceived thoughts on this topic and what we can expect from the incoming years.


\section{Constraining convection}

 Stars like the Sun have an outer convective zone where the energy is transported by convection motions. We have seen that the turbulence in the convection zone is responsible for the excitation of acoustic modes. It also plays a crucial role in the stellar magnetic activity. In a star, depending on the effective temperature and its evolutionary stage, the depth of the convective zone can vary. For { main-sequence} stars of spectral type F ($T_{\rm eff}$ between 6000~K and 7000~K), the convection zone is much thinner than in the Sun and can be as { shallow} as only a few percent of the stellar radius for the hottest stars, while in the Sun the convective zone represents $\sim$30\% of the solar radius. Stars cooler than the Sun, of spectral types K and M, 
have deeper convective envelopes than the Sun, and the coolest M-type dwarfs are even fully convective. 
When a solar-like star evolves and becomes a red giant, its core contracts and its outer envelope expands and the convective zone represents more than 50\% of the stellar radius. 
One way of determining the depth of the convection zone is to model the star and try to match the spectroscopic parameters ($T_{\rm eff}$, Fe/H...) and the frequencies of acoustic modes as seen in Chapter 3.2. The best fit model that is obtained provides the complete structure of the star and in particular the depth of the convective zone \citep[e.g.][]{2012ApJ...749..152M}. Of course, we have to remember that this model is obtained with our current knowledge of the physics in stellar evolution, which is not yet complete.

We can also study the pattern of the acoustic-mode frequencies, which allows us to define the mean large frequency separation and the small separation. However, the repetition of the pattern is not completely perfect. 
The transition from the radiative zone to the convection zone, which is a sharp change 
in the thermal gradient, affect the modes and thus their frquencies.
By studying the small deviation of the mode frequencies from the regular pattern, we observe an oscillation whose period is related to the size of the convective zone. We call it the ``acoustic glitches''. 
This method has been applied to the Sun \citep{JCDGou1991} and also to some solar-like stars \citep{2014ApJ...782...18M}. For a few tens of stars, the determination of the depth of the convective zone with stellar models and from the analysis of the acoustic glitches show a rather good agreement. 


Finally, one of the 
manifestations 
 of the convection is the granulation observed on the stellar surface where hot plasma rises towards the stellar surface and after cooling down descends back towards the interior. This phenomenon can be studied in the power spectrum computed with photometric observations for instance. We call it the stellar background, on top of which are observed the acoustic modes of the star. By fitting the stellar background { following the formulation described by \citet[][]{1985ESASP.235..199H}}, we can measure the time scale of the granulation, $\tau_{\rm gran}$:
 
 \begin{equation}
P_{\rm H}(\nu)=\frac{4\sigma^2\tau_{\rm gran}}{1+(2\pi\nu\tau_{\rm gran})^\alpha}\ ,
\end{equation}
{ where} $P_{\rm H}(\nu)$ is the total power at frequency $\nu$, $\sigma$ is the characteristic amplitude of the granulation and $\alpha$ is a positive parameter characterizing the slope of the decay. 
 
\begin{figure}[!htbp]
\begin{center}
\includegraphics[width=13cm, trim=1cm 0.8cm 0 0.8cm]{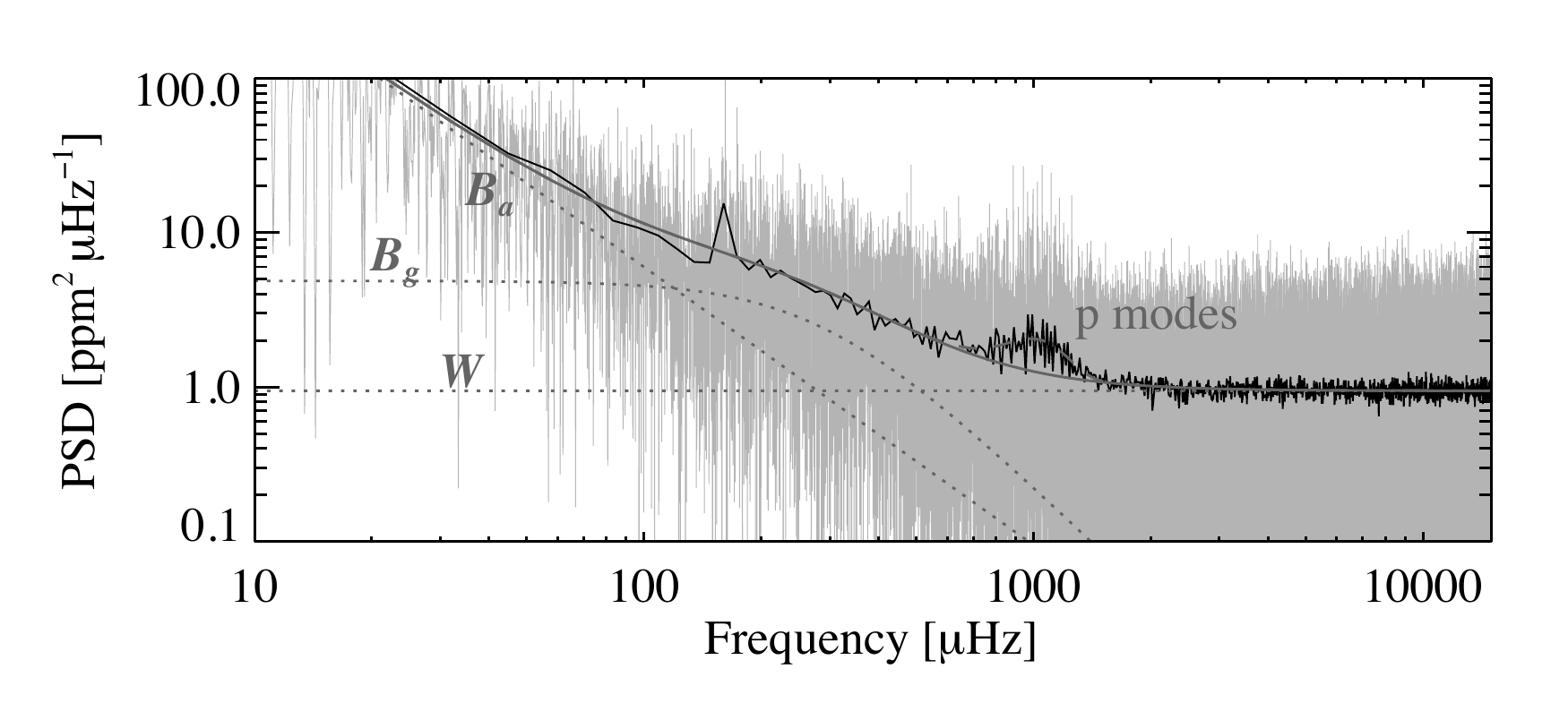}

\caption{Power spectrum density of the solar-like star HD~169392 (grey) and smoothed one (black)
observed by the CoRoT satellite. The background is represented by the white noise (W), the granulation ($B_g$) and a power law to reproduce the magnetic activity ($B_a$). Extracted from \citet{2013A&A...549A..12M}.}
\label{Background}
\end{center}
\end{figure}

 { First applied to the Sun, it has then been tested on a larger number of solar-like stars and red giants \citep[e.g.][]{2011ApJ...741..119M,2013A&A...559A..40S} showing the validity of the model.} An example is shown in Figure~\ref{Background}. The analysis of almost 1000 red giants showed that this time scale is inversely proportional to the mean large separation and to $\nu_{\rm max}$ \citep{2011ApJ...741..119M}. The scaling relations seen in Chapter~4.1 allow us to relate these two global parameters with the surface gravity of the star. Thus there is { an anticorrelation} between the granulation time scale and logg. The granulation time scale can also be studied in the time domain by analising the standard deviation of the light curve { on time scales shorter than 8 hours}. This quantity, called ``flicker'' has been studied by \citet{2013Natur.500..427B} and its relation with the granulation has been showed by \citet{2014ApJ...781..124C}.

\section{Constraining rotation}


The transport of angular momentum is very important to understand the { rotation history of stars. When they are born, their surface velocities depend on the conditions of the protostellar disk, and hence, a variety of rotation periods are observed at the beginning of the main sequence}. At this evolutionary stage, the existence of two different regimes delimited by the so-called Kraft break has been determined (Kraft 1967). Stars with masses larger than 1.3$M_{\odot}$ have such thin convective zones that the generation of magnetic winds does not happen, thus the rotation periods of theses stars barely evolve on the main sequence.  Stars below this mass limit loose angular momentum { very rapidly} via the magnetised stellar winds. The loss laws are of the form $dJ/d\omega$ \citep{1988ApJ...333..236K,1997ApJ...480..303K}, which means that fast rotators slow down more than slow rotators. This leads to a very narrow range of rotation periods at the end of the main sequence{   allowing the determination of gyrochronology relations \citep[e.g.][]{2003ApJ...586..464B}}. Finally, when the stars leave the main sequence, because of the expansion of the envelope, both regimes of stars undergo a spin down
of their envelope.
Models on the transport of angular momentum have been developed to understand the coupling that takes place in these different stages and regimes of the stars  \citep[e.g.][]{2013ApJ...776...67V,2014ApJ...780..159E}.

Most of the knowledge we have on stellar interiors is obtained from the calculation of stellar structure and evolutionary models (see Chapter 3.1). Those models are driven by many different physical mechanisms that are not perfectly understood. What is the effect of internal rotation in stars? First, centrifugal acceleration reduces local gravity and models of rotating stars behave as if they had lower masses. Second, rotation induces transport of chemical elements and angular momentum by the meridional circulation \citep{1925Obs....48...73E} and shear mixing \citep[e.g.][]{1992A&A...265..115Z,1998A&A...334.1000M,2004A&A...425..229M}, which modifies the global evolution of a star \citep{2009pfer.book.....M,2010A&A...509A..72E,2013A&A...549A..74M}.

Surface abundances of light elements can be used as indirect diagnostics of the efficiency of internal rotational mixing \citep{2010IAUS..268..365T} including, in particular, the use of lithium depletion \citep{2012A&A...539A..70E}. However, it is much more effective to directly compare the rotation rate inferred from the models with the observed one. This has been made possible thanks to the determination of the internal 
rotation profile obtained with seismic techniques applied to the Sun and other more distant stars.

In the case of the Sun, we have  thousands of high-degree p modes available allowing us to infer a very precise 2-D rotation map 
in the outer convective zone (see Fig.6 in Chapter 4.2). However, the number of acoustic modes probing the rotation rate in the radiative regions towards the center of the Sun reduces drastically and only low-degree modes provide reliable information in these deep layers. { \citet{ChaSek2004} studied the potential of low-degree p modes to measure rotation gradients between 0.1 and 0.3$R_{\odot}$. They concluded that changes of at least 110\,nHz would be needed to detect a gradient in the core rotation.} Indeed, the rotation profile in the Sun is only well constrained using p modes down to around $\sim$\,0.2 $R_\odot$ \citep{CouGar2003,2008SoPh..251..119G,2013SoPh..287...43E}.  To go beyond this limit it is necessary to measure mixed modes such as the $l$=2 and $l$=3 modes with $n$=1 \citep[see for example Fig.2 in][]{2004SoPh..220..269G} or individual gravity modes \citep{2007ApJ...668..594M,2008A&A...484..517M}. Unfortunately, those modes have tiny amplitudes at the surface of the Sun because they are evanescent in the convective region which extends for the outer 30$\%$ of the solar radius \citep{2010A&ARv..18..197A}. However, by measuring the asymptotic spacing of dipole gravity modes in the Sun --in a similar way to what is now commonly done in the analysis of red giants (see Chapter 4.1)-- it has been possible to determine that the core of the Sun below 0.2 $R_\odot$ could rotate faster than the rest of the radiative region \citep{2007Sci...316.1591G,2008AN....329..476G}. However, it would be necessary to measure individual splittings of those gravity modes to perform a proper inversion of the deepest solar layers to confirm this result.

In the stellar case the situation is different. For main-sequence solar-like stars, there are only a { few dozens of} low-degree p modes that can be observed. Their propagation cavities are very similar and we can only obtain an average rotation rate of the interior \citep{2013PNAS..11013267G} { weighted towards the surface}. Moreover, due to the width of the modes (inversely proportional to their lifetimes, see Chapter 2.2) and the correlation between rotational splitting and inclination angle, the extraction of reliable splittings is only possible for { stars with} high inclination angles { and} long-lifetime modes  \citep[e.g.][]{2006MNRAS.369.1281B}. { In any case, we can hope that in young F-stars with shallow convective envelopes, pure g modes could be observed as they will not be completely damped in this zone allowing to have a direct view of the conditions in the deep interior.}

The situation drastically changes when  stars are more evolved. In the sub giant and early red giant branch, when the hydrogen in the core is exhausted and stars only burn hydrogen in a shell, there are a few measurable mixed modes  that are very sensitive to the deepest radiative interior, in particular they provide invaluable information about the internal rotation \citep{2012ApJ...756...19D,2014A&A...564A..27D}. In this case, combining the information carried out by  acoustic and mixed modes, we can obtain the rotation at two distinct radii: one weighted towards the inner radiative interior and the other one weighted towards the external convective envelope.

Finally, in the case of red giants (burning H in a shell or after the ignition of the He in the core), the measurement of many  mixed modes  (typically more than 10) allows us to have a good determination of the rotation in the radiative core of these stars
\citep{2012Natur.481...55B,2013A&A...549A..75G}. 

The ensemble analysis of the rotation of hundred of red giants  \citep{2012A&A...548A..10M} confirms an increase in the rotation of the inner contracting radiative core compared to the expanding outer envelope but much less steep than what we could expect as a result of internal angular momentum distribution \citep{2000ApJ...540..489S}. This is a similar result to what we already knew from the analysis of the solar internal rotation profile \citep{2010ApJ...715.1539T}. This slower rotation rate in the core of red giants allows a better agreement with the results found in low-mass white dwarfs exhibiting slower rotation rates than what we expect also for conservation of angular momentum \citep{1999ApJ...516..349K}. 

These results confirm the necessity to evoke extra physical processes in the stellar evolution codes that could help to evacuate angular moment from the inner core towards the outer envelope \citep{2012A&A...544L...4E,2013A&A...555A..54C}. Two possible mechanism have been proposed. The first one invokes the existence of a fossil magnetic field in the stellar interior that were trapped during the first stages of the stellar evolution 
\citep{1998Natur.394..755G}. The second one implies the transport of angular momentum by internal gravity waves excited in the convective to radiative transition region \citep{2005A&A...440..981T}.

To complement the information on stellar internal differential rotation, the long and uninterrupted photometric time series required for asteroseismic studies are excellent to measure any modulation in the photometry induced by stellar spots crossing the visible stellar disk \citep{2013ASPC..479..129G,2013ASSP...31..215L}, which are proportional to the surface rotation rate at the active latitudes. The information on the surface rotation can be extracted by performing detailed spot modelling \citep{2009A&A...506..245M,2010A&A...518A..53M,2012A&A...543A.146F} or by more general treatments using automatic analysis of the cross correlation of the temporal signal \citep{2013ApJ...775L..11M}, the analysis of the low-frequency part of the periodogram \citep{2013A&A...560A...4R,2013A&A...557L..10N}, or by using time-frequency analysis coupled to the other two methods \citep{2014A&A...562A.124M,2014A&A...572A..34G}. The determination of surface rotation rates { allows} a natural link with classical analysis based on surface rotation using spectroscopy. They impose { strong} constraints on the rotation of the external layers for sub giants and giants while providing also information about the existence of surface magnetic field at these stages of evolution. { Asteroseismology on the other hand provides information on the rotation profile in deeper layers of the stars, which is crucial to understand the transport of angular momentum.}

\section{Constraining magnetic fields}



Different techniques are used to measure magnetic fields and magnetic activity in stars. Classical observations are based on the measurement of optical, ultra-violet, or X-ray emission from the magnetically heated gas. Starspots and flares that are associated with magnetic fields are also used to infer their magnetic activity.

Spectroscopic surveys such as the one at the Mount Wilson Observatory \citep{1978ApJ...226..379W} consist of measuring changes in spectroscopic lines (such as CaHK, H$\alpha$...) due to the magnetic field present in the stars. These observations that are done for several years in a large number of stars have allowed us to observe and discover a broad variety of magnetic activity behaviours during the last decades for stars with different fundamental properties (rotation, mass, spectral type, evolutionary stage...). These studies suggested a relation between the length of the magnetic cycles and the surface rotation periods \citep{2002AN....323..357S,2007ApJ...657..486B}. This also points out the existence of two different branches called the ``inactive'' and the ``active'' branch.  No clear explanation on the differences between the two branches has been given so far but the most common one implies a different location of the dynamo process more or less deeper in the stars.

Theory shows that the rotation and the magnetic cycle periods are related to the Rossby number Ro=$\tau_c/P_{\rm rot}$, where $\tau_c$ is the convective turnover time and $P_{\rm rot}$ is the surface rotation period.
 In order to understand the detailed mechanisms responsible for the existence of magnetic activity cycles it is crucial to have a precise knowledge of the internal stellar structure and rotation. From so many decades of observations of the Sun, we believe that for a star like the Sun, with an outer convective zone, the magnetic activity results from a dynamo process where rotation, convection, and magnetic fields are interacting all the time. From this picture, we would then expect fast rotators to have shorter magnetic activity cycles that can be observed with missions such as {\it Kepler} or PLATO. Finally, spectroscopic observations by \citet{2006A&A...446..267R} suggest that faster rotating F-type stars with masses larger than 1.2$M_{\odot}$  have a stronger differential rotation, which is an important ingredient in the magnetic activity and make these stars interesting study cases.

In the last few years, cycle lengths shorter than a few years have been observed with spectroscopic and spectropolarimetric observations. By analysing the electromagnetic spectrum, we can study the variation of  some absorption lines that are sensitive to the magnetic field. A classical proxy of the magnetic activity is the S-index, based on 
Ca{\sc ii} H and K lines, that is sensitive to the magnetic field in the photosphere and the chromosphere. For the Sun, the average S-index is around 0.171. A good example of such spectroscopic study is the F-type star HD17051 (also known as 
$\iota$ Horologii) that was monitored in 
Ca{\sc ii} H and K with SMARTS 1.5m telescope for five years. This study suggested a cycle period of 1.6 year for a surface rotation of 8 days \citep{2010ApJ...723L.213M}. However longer datasets did not show another cycle but a chaotic variation of the S-index \citep{2013A&A...553L...6S}. 

Spectropolarimetric observations consist of observing the light spectrum and also measuring the polarisation of the light  over the whole spectral range, which allows us to study large-scale magnetic fields in the stars. Because the star rotates, at different rotational phases, we can observe different distributions of the magnetic field. Polarised spectra are collected by spectropolarimeters such as ESPaDOnS at the CFHT (Hawai), NARVAL on TBL (France), and HARPSpol  (Chile).  Some very interesting results have been obtained for solar-like stars the last few years.  
\citet{2009MNRAS.398.1383F} 
 detected 
a change of polarity in the magnetic field of the F-type star $\tau$ 
 Boo, the star with the shortest cycle period observed.  The Bcool project detected magnetic field in 67 stars through theyr spectropolarimetric survey of 170 stars. This study shows that magnetic mapping can be done for a variety of moderately active stars \citep[see][for more details]{2014MNRAS.444.3517M}.

With the long and uninterrupted asteroseismic measurements collected by the CoRoT and {\it Kepler} missions, 
 we have a unique database of lightcurves to study activity cycles in other stars for which we can probe the internal layers as well. { We already talked} about using the time-frequency analysis to measure the surface rotation period. As this measurement is linked to the presence of spots and thus magnetic activity this time-frequency analysis detects signatures of magnetic activity cycles. \cite{2014A&A...562A.124M} analysed 22 F-type stars { observed for 1400 days} with a time-frequency method leading to the detection of magnetic activity signature for two stars. This study should be extended to more stars with different spectral types and rotation periods to have more statistics on the existence of cycle in the stars and check the accuracy of the rotation-activity-age relation.

Seismology also allows us to look for hints of magnetic activity. Indeed for the Sun, we know that when the magnetic activity increases, the acoustic modes have smaller amplitudes while their frequencies shift towards higher values \citep{2009A&A...504L...1S}. This can be explained by the fact that there is a change in the effective temperature affecting the sound speed \citep{1991ApJ...370..752G}  and thus the frequencies of the modes. Do we see the same pattern in other stars? \citet{2010Sci...329.1032G} studied the CoRoT target, HD~49933, which is an F-type star with a rotation period of 3.4 days. Similarly to what is seen in the Sun, they detected the same anti-correlation in the acoustic-mode parameters suggesting that the star was going through a magnetic cycle \citep[see Figure~1 of][]{2010Sci...329.1032G}, which was confirmed later by spectroscopic observations. Furthermore, a deeper analysis of this star showed that the shift in frequency is larger for the higher frequency modes \citep{2011A&A...530A.127S}. This is also what we observe in the Sun and it suggests that the magnetic changes happen closer to the surface as the high-frequency modes are more sensitive to the outer layers.
{ In the case of the existence of a magnetic field in the stellar deep interior, for example of fossil origin, modes propagating in these regions would be split in different components separated by a magnetic splitting that would be proportional to the magnetic field strength \citep{GooTho1992,2007MNRAS.377..453R}. However, a first analysis of the fine structure of mixed modes in 300 red giant stars performed by \citet{2012A&A...548A..10M}, showed that all of the measured mixed modes could be interpreted as rotationally split. If a magnetic field exists in those stars, it should be confined very deep or its strength should be too small to be measured with the current observations.  
}


\section{Seismology of rotating stars} 

Rotation has a direct impact on the oscillation modes and their frequencies. Some of those effects have already been discussed in chapter 4.2 but they are only valid as long as the rotation is slow. Before going further, we should clarify what is a ``slow rotation''. The rotation is slow when the effects of both the Coriolis force and the centrifugal distortion are small. By denoting $\Omega$ the rotation rate, this leads to the following conditions: $\Omega \ll \omega$ and  $\Omega \ll \Omega_\mathrm{K}$, where $\omega$ is the oscillation frequency, while $\Omega_K = \sqrt{GM/R^3}$ is the Keplerian break-up rotation rate, with $G$, $M$, $R$ the gravitational constant, the mass and equatorial radius of the star, respectively. When a star spins at its Keplerian break-up rotation rate, the centrifugal force at the surface, along the equator, is exactly balanced by the gravity; hence it corresponds to the maximal rotation rate above which the star would break up.

When these two conditions ($\Omega \ll \omega$ and $\Omega \ll \Omega_\mathrm{K}$) are fulfilled, it is possible to treat the effects of the rotation as a small perturbation. In such a case, the first step is to solve the oscillation equations neglecting the rotation. We then obtain a set of eigenmodes characterized by their quantum number $(n,l,m)$ associated with eigenfrequencies $\omega_{n,l}$ which are independent on $m$. The first-order correction in $\Omega$ does not affect the modes, only the frequencies are modified. We can identify two contributions to this first-order term: 1) the waves are advected by the 
rotation, then prograde (retrograde, resp.) 
modes have slightly higher (lower, resp.) frequencies than zonal ($m=0$) modes, which stay unaffected. Prograde (retrograde) modes are modes with $m<0$ ($m>0$), i.e. modes propagating around the axis in the same direction as the rotation (the opposite direction).
2) Coriolis force also modifies the frequency, but this is generally very weak for p modes, 
 only g modes are affected. 

If we need to take the second-order effects ($O(\Omega^2)$) into account, the centrifugal distortion must be considered, and both the frequencies and the spatial distributions of modes are perturbed. Methods which take into account the first-order effects of rotation were developed in \citet{Cowling49,Ledoux51}, second-order corrections were proposed 
for example by \citet{Saio81,DziembowskiGoode92,Suarez06}.

Perturbative approaches are the easiest way to treat slow rotation. However, for g modes --which have long periods-- the condition $\Omega \ll \omega$ is generally not fulfilled. 
{More general methods should then be used in such cases. We can see for example the work of \citet{1987AcA....37..313D}. However the most classical approach, developed and commonly used in geophysics \citep{Eckart60}, is the so-called traditional approximation.} Within this approximation, the oscillation equations are rewritten, leading to a problem that is also separable in $r$ (radius) and $\theta$ (colatitude). However, the horizontal dependency of modes is no more described with spherical harmonics, as in the non rotating case, but with Hough functions \citep[e.g.][]{Unno89}. This traditional approximation has been applied to compute g modes in various stars 
\citep[see, for example][]{Berthomieu78,Lee92,Townsend97,Savonije05}.

In practice, the second assumption ($\Omega \ll \Omega_\mathrm{K}$) is no longer valid for $\Omega \ge 0.1\Omega_\mathrm{K}$ \citep{Reese06,Ballot10}. Above this limit, perturbative approaches fail in recovering correct mode frequencies. For a typical A-type star, this would correspond to an equatorial velocity $v\ge 50\ \mathrm{km s^{-1}}$. Unfortunately such a situation is very common for intermediate- and high-mass main-sequence stars \citep[e.g.][]{Royer07}.
As a consequence, rotation may strongly impact the frequencies of p modes, especially in $\delta$ Scuti and $\beta$ Cephei stars, and g modes, especially in $\gamma$ Doradus and slowly pulsating B-type (SPB) stars. Complete computations of oscillations must then be performed to solve this problem. 

The first challenge is to model a distorted star. Taking advantage of the spherical symmetry, non-rotating stars are modelled as 1-D objects (the structure depending only on the radius $r$). However, rapidly rotating stars must be considered as 2-D objects, for which physical quantities are function of $r$ and $\theta$. 
New generations of more realistic stellar models are now emerging%
\citep[e.g.][]{Deupree11,Roxburgh06,MacGregor07,Rieutord13}. 

The second challenge is to solve the system of equations governing the oscillations in the presence of rotation. In the non-rotating case, the oscillation problem is separable with respect to the radial and horizontal variables:  the horizontal component of a mode is described with a spherical harmonic of a given degree $l$ (and azimuthal order $m$), thus, for a given $l$, we only have to solve a 1-D problem. This is not true for a rotating star: the problem is no longer separable with respect to $r$ and $\theta$; as a consequence, one mode cannot be described by a unique spherical harmonic.

\begin{figure*}[!htbp]
\label{fig:modesrot}
\includegraphics[width=0.45\linewidth]{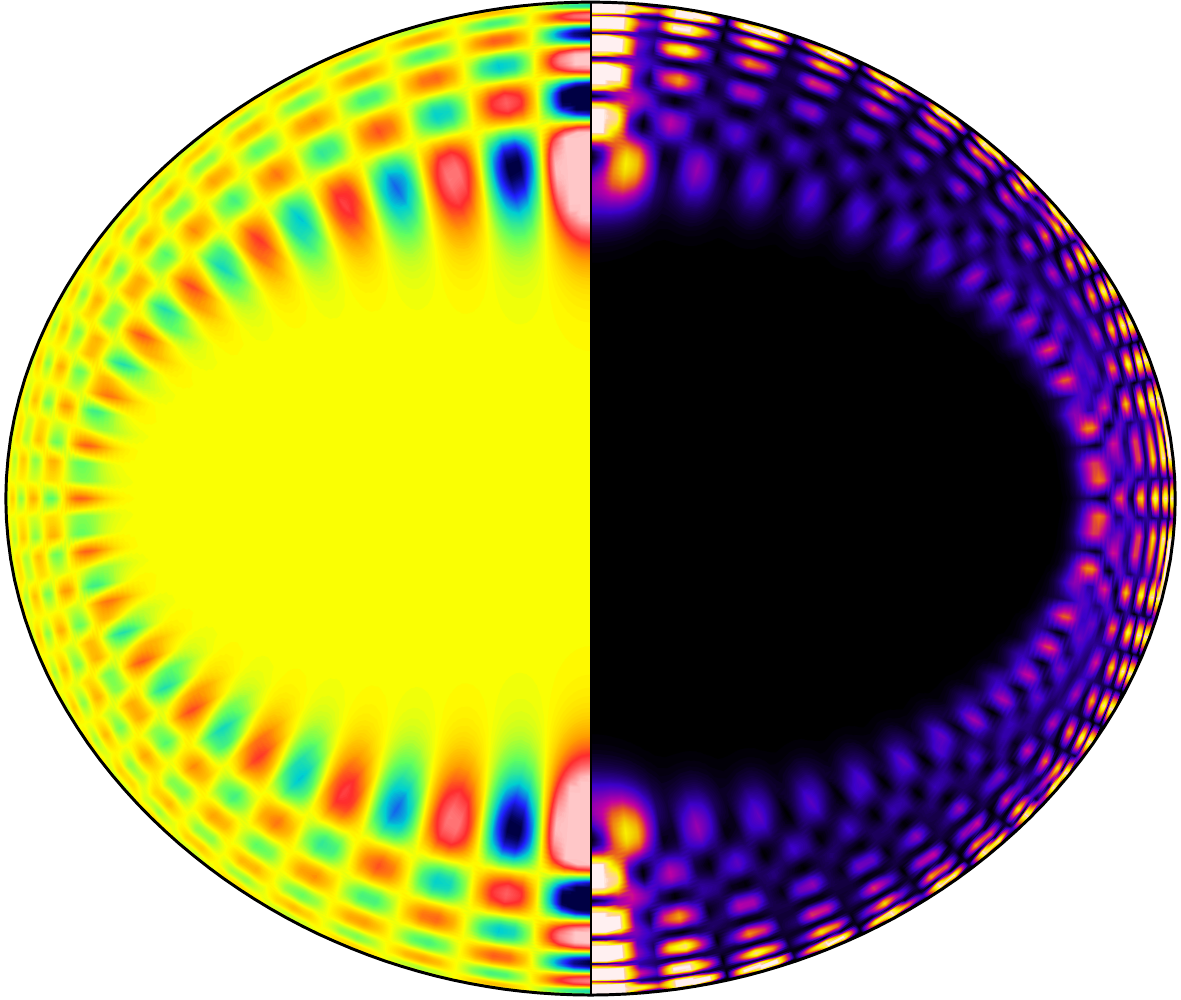}%
\includegraphics[width=0.45\linewidth]{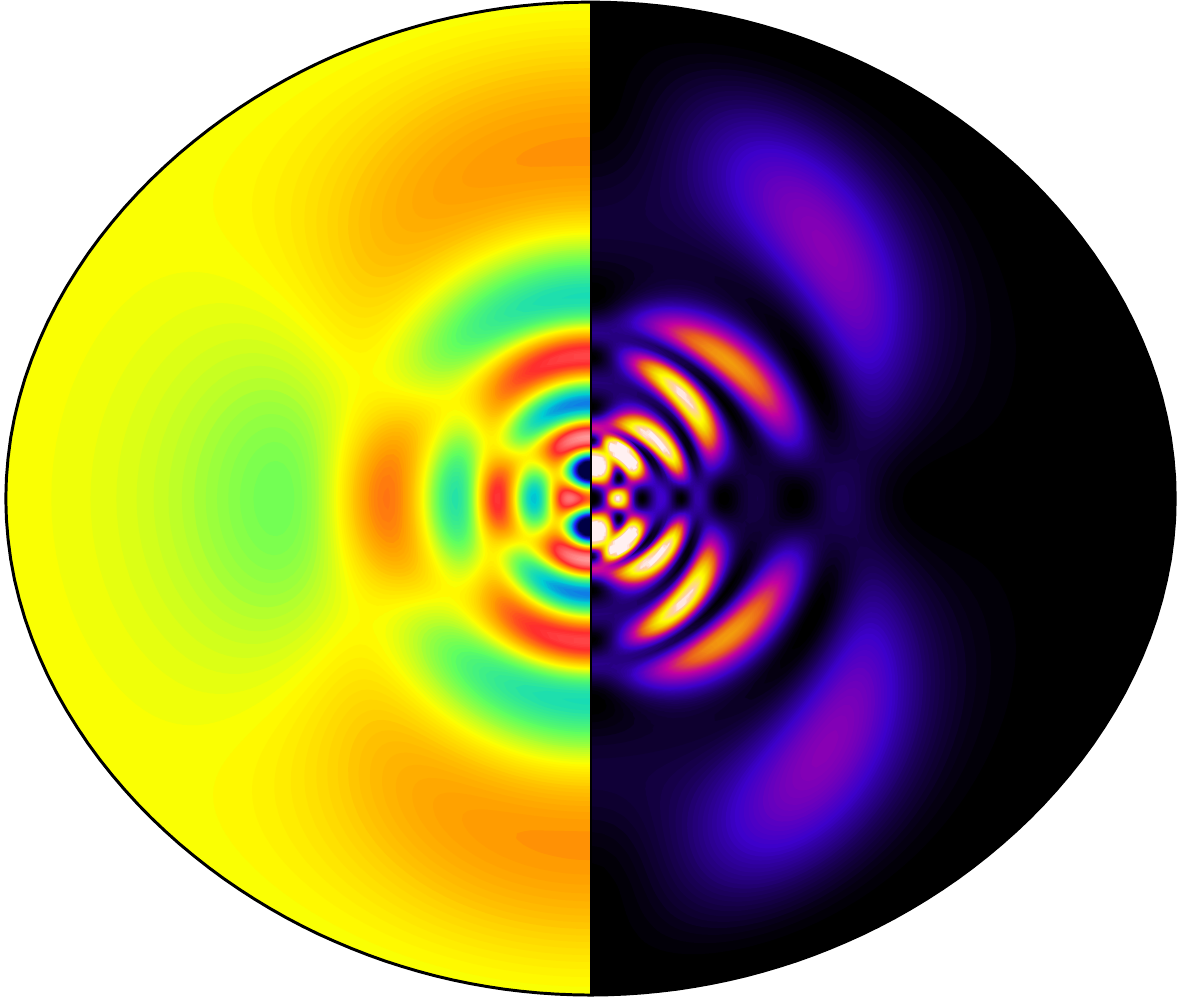}\\
\includegraphics[width=0.45\linewidth]{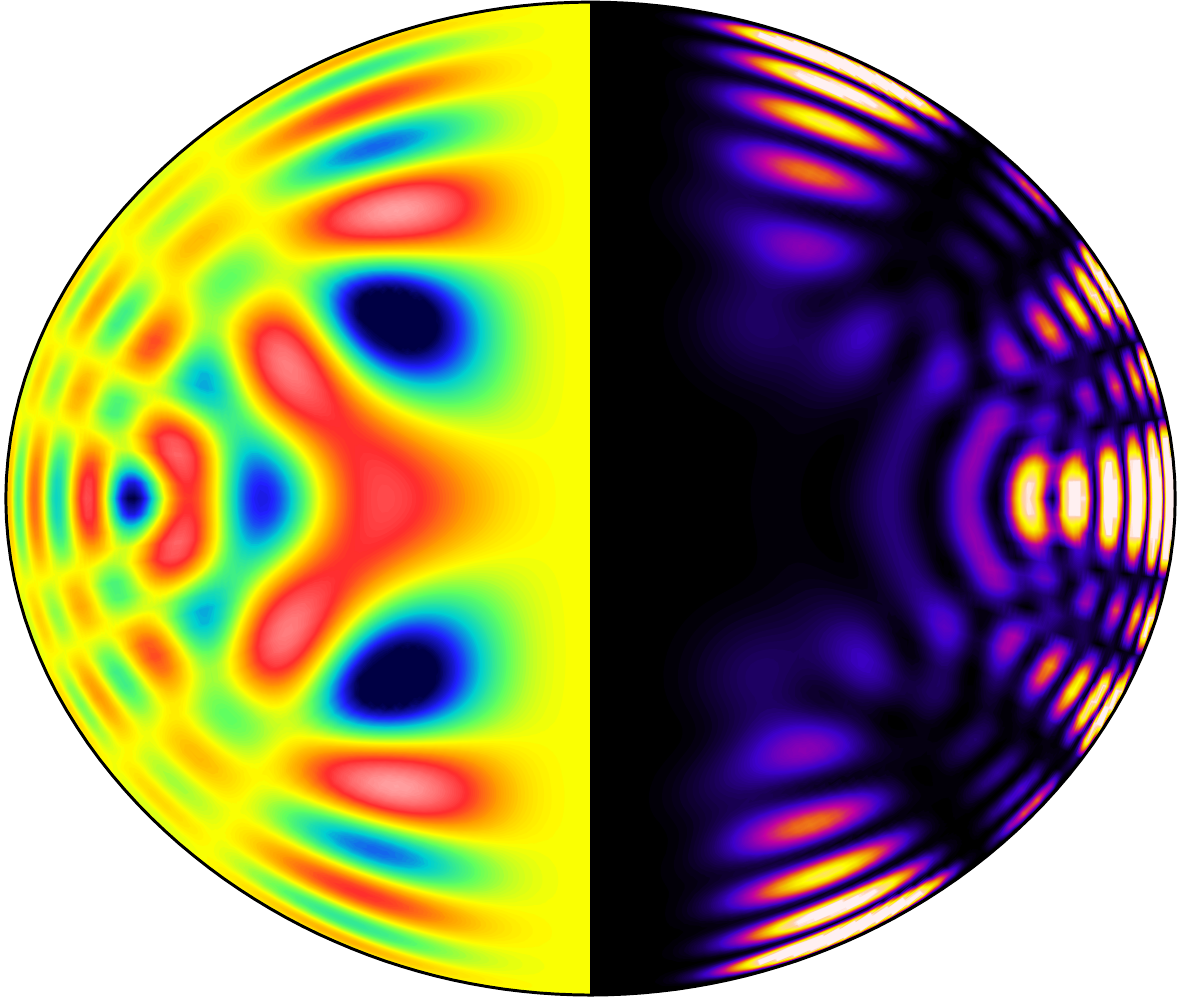}%
\includegraphics[width=0.45\linewidth]{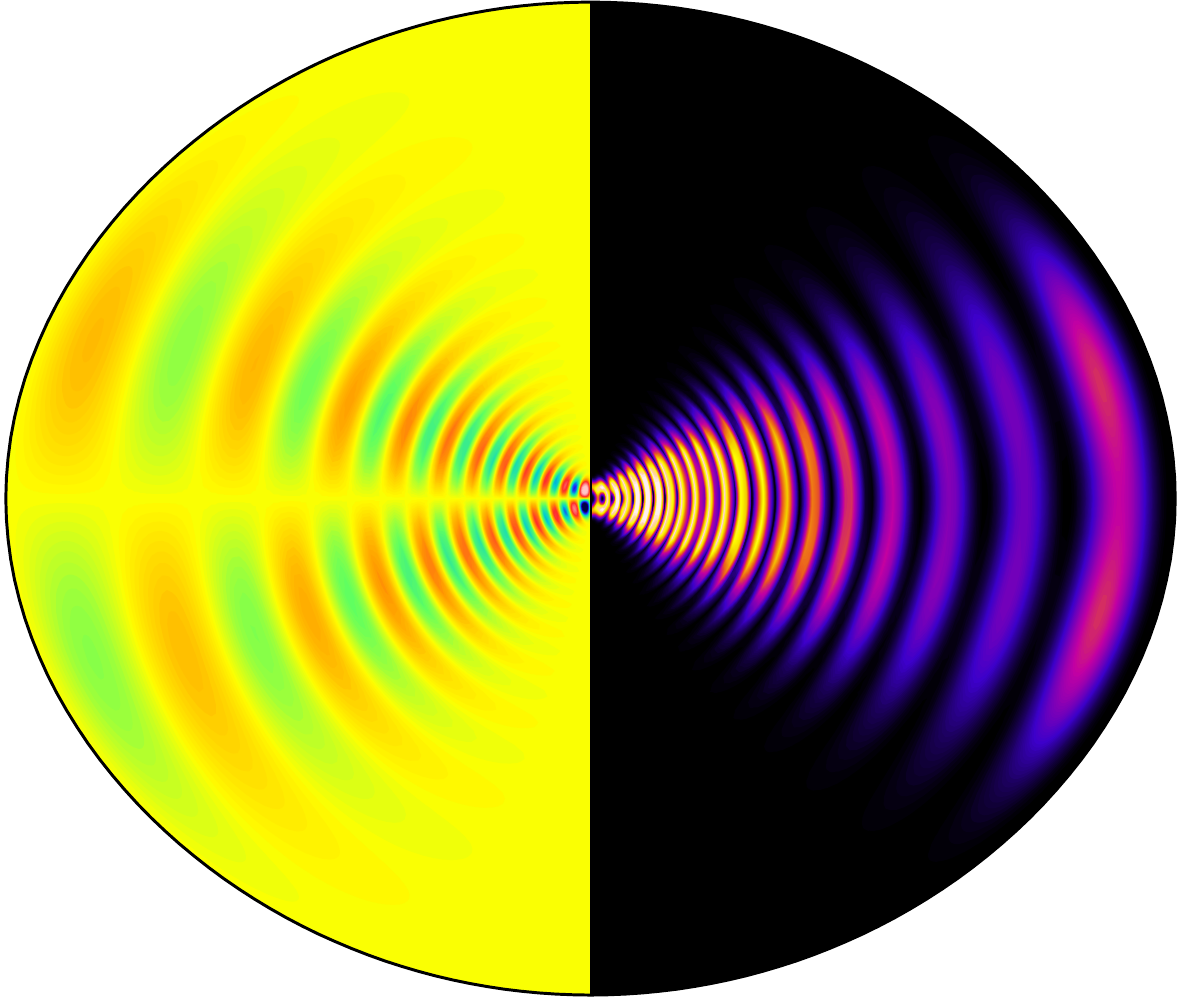}\\
\includegraphics[width=0.45\textwidth]{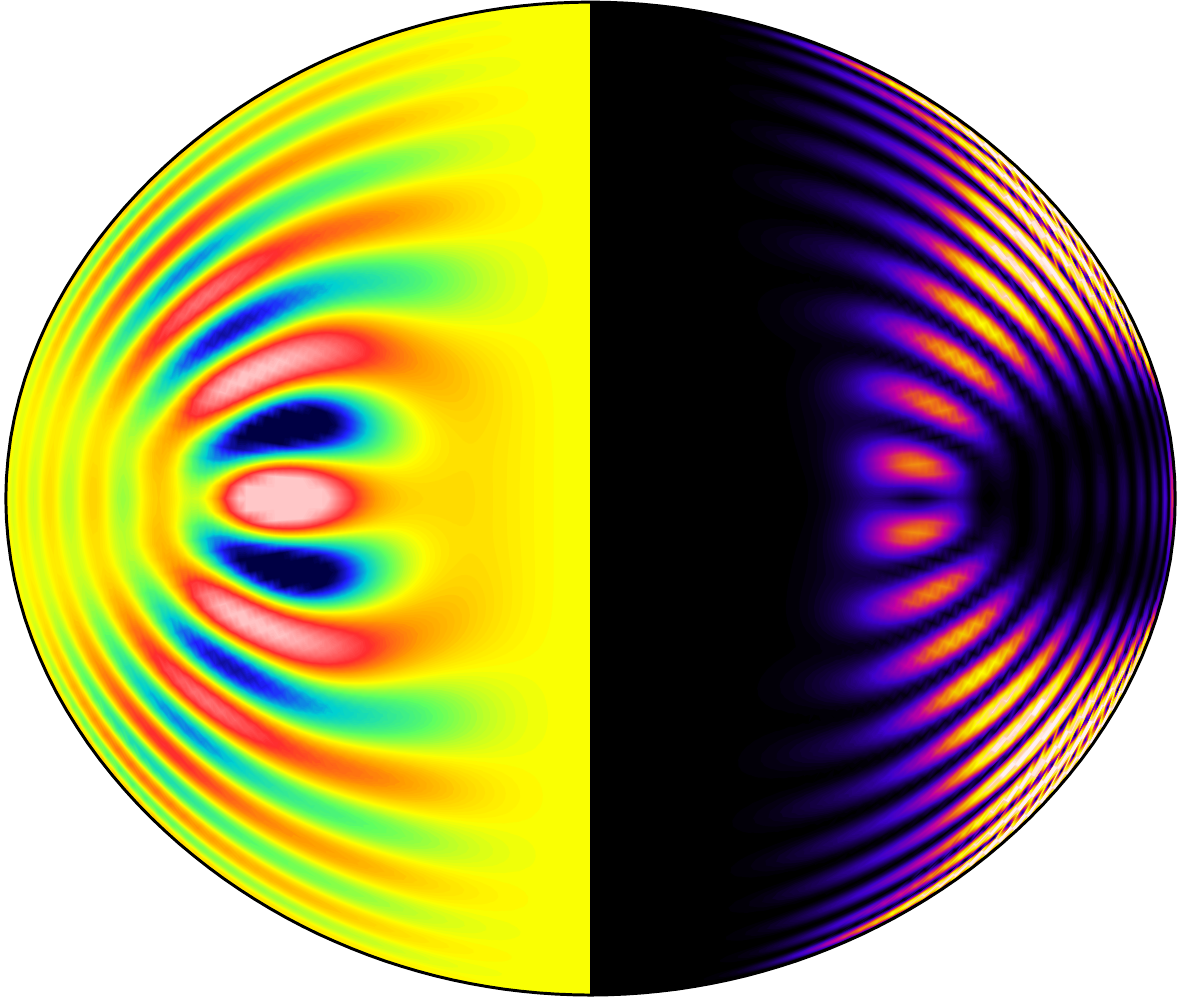}%
\includegraphics[width=0.45\linewidth]{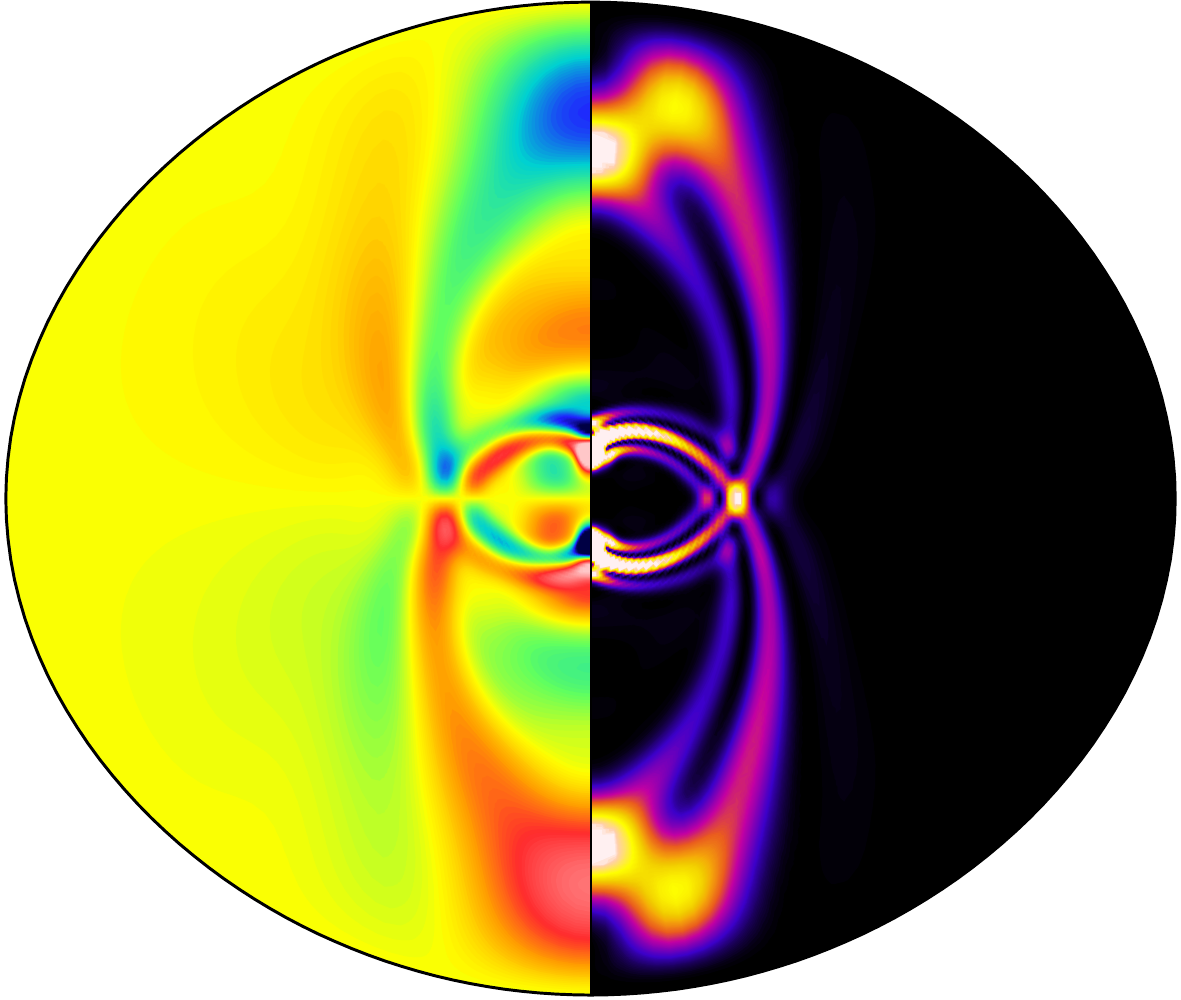}
\caption{Examples of modes in a rotating star ($\Omega=0.6\Omega_\mathrm{K}$). First column: three representative p modes (top to bottom: whispering gallery mode, chaotic mode, island mode); Second column: three representative g modes (top to bottom: Mode in the super-inertial domain ($\omega>2\Omega$) not too affected by rotation, mode in the sub-inertial domain ($\omega<2\Omega$) affected by a forbidden region in the polar region, a rosette mode in the super-inertial domain. In each panel: on the left-hand side, the variable $p' / \sqrt{\rho}$ is shown in a meridional plane ($p'$ is the pressure fluctuation and $\rho$ the density); on the right-hand side, the kinetic energy of the mode is represented in a meridional plane.}
\end{figure*}

Two-dimensional oscillation codes have been developed to solve this problem 
\citep[e.g.][]{Clement84, Lee95, Savonije95, Reese06, Ouazzani11}. More details and discussions on technical aspects can be found in \citet{Ballot13}. Recent investigations with accurate codes, such as the TOP code \citep{Reese06,Reese09}, allowed us to better understand how the oscillation spectrum is reorganised in fast spinning stars. To illustrate this, we compute oscillations in a simplified stellar structure consisting in a polytropic model with a polytropic index $\mu=3$
(i.e. a model  where pressure $p$ and density $\rho$ are linked through the relation $p \propto \rho^{1+1/\mu}$), that mimics a radiative star. 
Typical modes are shown in Fig.~\ref{fig:modesrot} for a uniform rotation rate $\Omega=0.6\Omega_\mathrm{K}$. 

If we consider p modes, we notice that they present very different spatial structures. 
{A ray theory for rapidly rotating stars is very helpful to understand these differences. We can notice a first attempt by \citet{1988IAUS..123..125P} in this direction. The recent work of \citet{Lignieres08,Lignieres09} allowed us to properly classify the modes.} For a given $m$, we can distinguish between 3 families of modes. 
First, the whispering gallery modes are similar to the modes found in non-rotating stars. At high rotation rates, they have a large number of latitudinal nodes. 
Second, we have the chaotic modes, so called because they are associated with chaotic trajectories in the ray dynamics. The modes by themselves do not have any chaotic behaviour. They are characterized by irregular nodal lines and they propagate down to the stellar centre. 
Third, the island modes are associated with stable periodic trajectories in the ray dynamics. The energy of the mode is concentrated around this stable trajectory. We can count the number of nodes along this trajectory (denoted $\tilde n$) and in the transversal direction (denoted $\tilde l$). For a given $m$, these two numbers ($\tilde n$,$\tilde l$) fully characterize an island mode. For example, the island mode plotted in Fig.~\ref{fig:modesrot} is characterized by $\tilde l=0$ and $\tilde n$=20 (and $m=3$). Island modes correspond to the low-degree modes of the non-rotating case.

The p-mode spectrum of a rotating star is then the superimposition of subspectra corresponding to these different families. Thus, the global spectrum seems more complex, and the regularity of the spectrum, described at $\Omega=0$ by the Tassoul's relation \citep{Tassoul80}, $\omega_{n,l} = \Delta_0 ( n + l/2 + \alpha )$, is apparently lost. However new regularities appear inside the families. The regularities of the island modes have been extensively studied numerically by \citet{Lignieres06,Reese08,Reese09} and theoretical by \citet{Pasek11,Pasek12}. It has been shown that, for island modes: 
$ \omega_{\tilde n,\tilde l,m} = 
  \tilde n\tilde\Delta_{n} + 
  \tilde l \tilde \Delta_l + 
  \left|m\right| \tilde \Delta_m - m\Omega +
  \tilde\alpha^{\pm}$ 
where $\tilde\Delta_n$, $\tilde\Delta_l$, $\tilde\Delta_m$, and $\tilde\alpha^{\pm}$ are parameters that are related to the stellar structure. 
Recovering regularities is a crucial question because regularities \textbf{are} very helpful in identifying modes in observations, 
and mode identification is a prerequisite for any sensible inference on internal structure and dynamics.

For g modes, the spatial structure will first depend on its frequency $\omega$, expressed in the co-rotating frame (not the observer's one). Generally, when a mode stays in the super-inertial regime ($\omega > 2 \Omega$), its structure stays similar to the one of a g mode in a non-rotating star and is only slightly deformed. However, for a mode in the sub-inertial domain ($\omega < 2 \Omega$) a more visible change is observed due to the appearance of a polar, approximately conical, forbidden region where waves cannot propagate \citep[see][]{Dintrans00}. The critical angle defining this cone is
close to $\arccos(\omega/2\Omega)$. Modes are then more and more confined around the equator plane when $\omega/\Omega$ decreases.
A new family of modes has been discovered recently, the so-called rosette modes \citep{Ballot12}. These modes have been found in the super-inertial regime and exhibit a drastic change in their structure. For these modes, the kinetic energy is focused around a set of loops, forming a ``rosette'' pattern. They appear for rotation rates as low as $\Omega\approx 0.1 \Omega_\mathrm{K}$. The formation of the rosette modes can be understood with two different approaches: 1) Similarly to the island p modes, they can be associated to stable periodic trajectories in the ray dynamics \citep{Ballot12}; 2) Their formation can also be explained as an interaction among g modes having close frequencies \citep{Takata13}.



\bibliography{rotation,BIBLIO,BIBLIO_jb} 


\end{document}